\documentclass[trackchanges]{aastex7}
\usepackage{graphicx}%
\usepackage{amsmath,amssymb,amsfonts}%
\usepackage{mathrsfs}%
\usepackage[title]{appendix}%
\usepackage{textcomp}%
\usepackage{listings}%
\newcommand{\etal}{\mbox{\rm{et al.}~~}}
\newcommand{\kms}{\mbox{km\,s$^{-1}$}}

\begin{document}

\title{Gaia's faintest stars}

\author[gname=Jeremy,sname=Mould]{Jeremy Mould}
\email{jmould@swin.edu.au} 
\affiliation{Centre of Astrophysics \& Supercomputing, Swinburne University,
Hawthorn, Vic 3122, Australia}
\affiliation{ARC Centre of Excellence for Dark Matter Particle Physics}

i






\begin{abstract}
In the {\it Brief History of Time} Stephen Hawking was pessimistic about astronomers detecting primordial black holes (PBHs). He would not be the only distinguished scientist to underestimate the extraordinary power of new technology. In a related area Albert Einstein published the equations for microlensing, but wrote off their practicality. Perhaps they meant ``during my lifetime.''
The amazing properties of PBHs, however, validate heroic efforts to detect them. If they exist, their niches in our current history of time include supplying dark matter to bind galaxies, offering a solution for the Hubble tension, and, as supermassive black holes, giving us quasars as far as the eye can see.
This Research Note describes a search for PBHs in the $Gaia$ archive. In spite of the high density of local dark matter, it was unsuccessful. Microlensing with the Rubin telescope is the tool at our disposal to open the asteroid window for PBH.
\end{abstract}








\section{Introduction}
 If they make up much of the Galaxy's dark matter (DM), the 
temperatures of subsolar mass primordial black holes (PBHs)
and the local density of DM  might suggest that some PBHs reside in
the $Gaia$ archive. Comparison of  PBHs (Mould 2025) in the Hertzsprung-Russell (HR) diagram  
with white dwarfs and neutron stars shows that, if they exist, they are
the most compact of compact objects. The utility
of reduced proper motion (RPM) in finding low luminosity stars
recommends that technique for testing this possibility.

A sample is presented in $\S$2, and the evolutionary state
of these stars is then discussed ($\S$3). Brown dwarfs account for
most of the stars, but there is also a possible role for PBHs.

\begin{figure}[h]
\includegraphics[angle=-0,width=\textwidth]{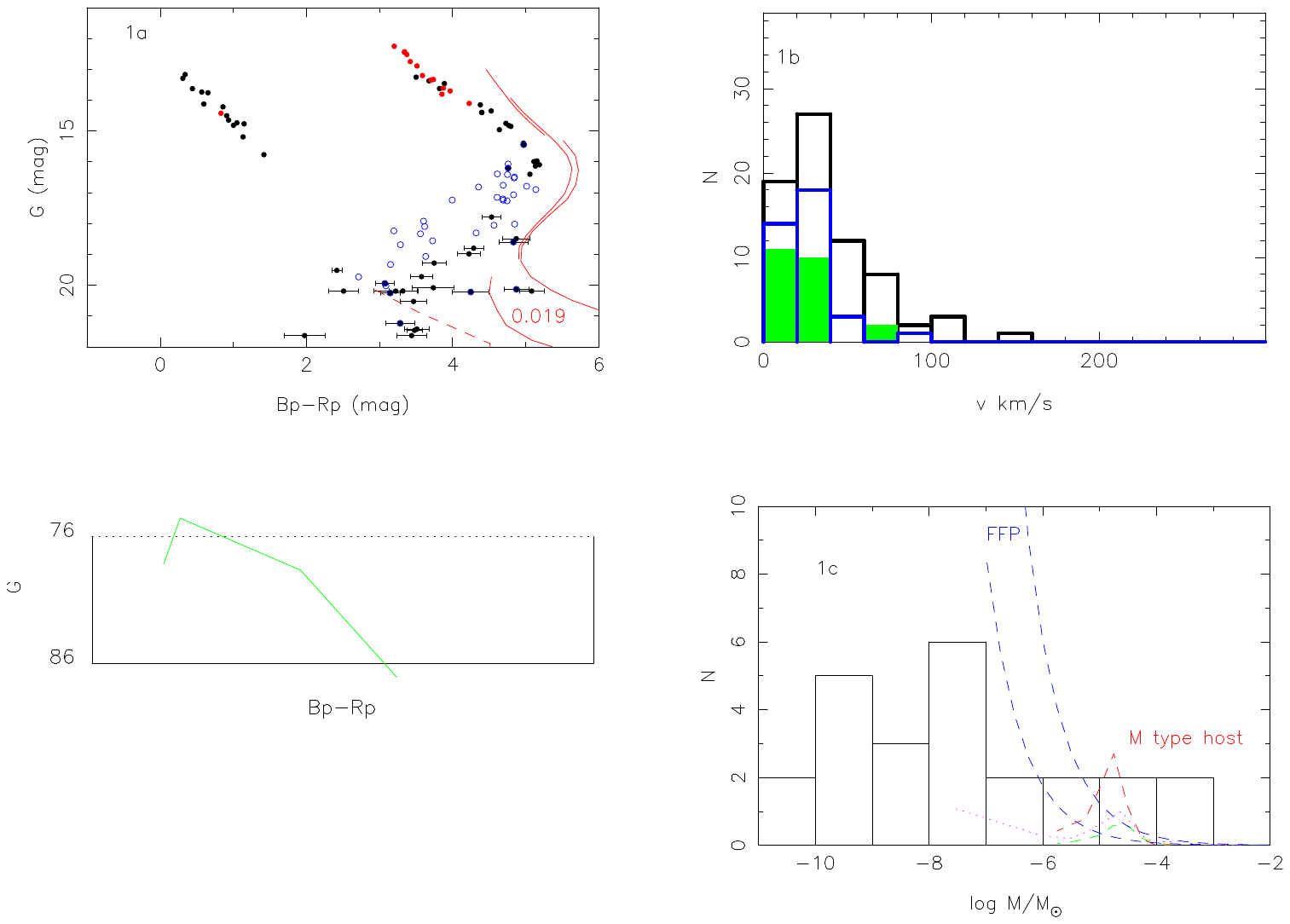}
\caption{(a) HR diagram for the sample. Stars in red have $Gaia$ radial
velocities. Brown dwarfs have been added
	(blue open circles, the UCD sample). 
	The red lines are Atmos2020 models (Phillips \etal 2020)
with ages exceeding 0.5 Gyrs and masses near 0.05 M$_\odot$; the branch
	labeled 0.019 is 0.019 M$_\odot$, and
	the dashed red line is a temperature sequence 
	at 0.1  Jupiter radius (R$_J$). A model for Jupiter mass and age 1 Myr
        is at G = 32 and Bp--Rp = 4.
	The lower extension to the figure
	shows the predicted G magnitudes of m8 to m11 PBH, m10 being the optically brightest. 
    (b) Tangential velocities of the sample (black). Stars fainter than 
	G = 17 mag (solid green). UCD
	stars from Sarro \etal (2022, blue). In spite of RPM selection these are disk kinematics.
    (c) The histogram is the mass distribution of solar system planets.
	The number of microlensing planet mass objects per host star
	is shown for M stars in red and G stars in green (Suzuki \etal 2016). The blue
	dashed lines bound the distribution of Free Floating Planets (FFPs)
	claimed by Sumi \etal (2023). The purple  dotted line is a theoretical
	FFP distribution by Coleman \& Derocco (2025).}
\end{figure}
\section{A Gaia archive sample}
I list\footnote{https://github.com/jrmould/Gaia} the $Gaia$ DR3 objects with g + 5 log $\mu~>$ 20 mag and parallax exceeding 100 milliarcsec (mas), where g is $Gaia$ magnitude and $\mu$ is
the proper motion in mas/yr. 
The absolute magnitude, G, assumes the distance is given
by the parallax, and the color and parallax uncertainties are in the
right hand columns. The latter is small enough that Lutz-Kelker (1973)
statistical corrections are not required.

Figure 1a is their HR diagram. At the bright end two sequences are very clear,
the main sequence of hydrogen burning stars and the white dwarf cooling
sequence. Stars fainter than G = 17.5 mag are something else. We have added
an ``ultra cool dwarf" (UCD) sample from Sarro \etal (2023). Figure 1b
shows the kinematics of both samples.  
The green
line connects the Hawking temperatures and luminosities of PBH from m11 (which I denote as
10$^{-11}$ M$_\odot$), through m9 to m8 
(10$^{-8}$ M$_\odot$). Bolometric corrections for PBHs were calculated
from the filter transmissions https://www.cosmos.esa.int/web/gaia/edr3-passbands and blackbody fluxes.

The color discrepancy in Figure 1a 
between the tracks in red and the data might be due to 
missing red opacities in the 1100--2400K temperature range of the UCDs.
The models were clearly labeled ``Vega", which is consistent with the $Gaia$ color zeropoint, and the chemical equilibrium variant was used. The non-equilibrium
models are 0.5 mag redder.
\section{Low luminosity extrema}
White dwarfs peter out at G = 16,
and observed neutron stars at L/L$_\odot$ = 10$^{-8}$. The 20 stars fainter than
this are truly low luminosity.
White dwarf and neutron star tracks are cooling sequences with Gyr and Myr respective timescales 
(Bottaro, Caputo, \& Fiorillo 2024). 
The former has an endpoint due to the age of the 
Galaxy. PBHs, on the other hand, are an evaporation sequence with an inverse relation between temperature and mass. The sole
candidate to date (Niikura \etal 2019) is at $\sim$10$^{-11}$ M$_\odot$, although unintended detections up to 10$^{-4}$ M$_\odot$ may have occurred in 
microlensing surveys (Scholtz \& Unwin 2019) 
towards the Galactic Center  (Mould 2025).
\subsection{Free Floating Planets}
Microlensing discoveries of planet mass objects are also shown in Figure 1c.
These objects are arbitrarily faint, but unless they are young, they
are also cold 
 and extremely faint (e.g. Jupiter is G $\approx$ 32). However, only 1 ppm of the stars in the $Gaia$ archive have G $<$ --2 mag (Mazzi \etal 2024). The solar neighborhood has too few young stars
to have been the original hosts of planets that are still warm enough
to be luminous, 
and young regions like Taurus or Orion are at least 150 pc away with a travel time of 5 Myrs at 30 \kms.

 
\subsection{Population II}
In population II low mass stars lie below the old disk main sequence. However, nothing in the HST main sequences of NGC 6397 and 47 Tuc (Richer \etal 2008, Heyl
\etal 2017), which reach the hydrogen burning limit, resembles the UCD sample in Figure 1.  Furthermore, everything in Figure 1 has disk kinematics.
\subsection{Compact UCDs }
The simplest explanation for the faintest stars
in Figure 1 is that they resemble the UCD sample,
but have smaller radii. Indeed, the models in Figure 1 have alternates with different equations of state, and these do have smaller radii. However, this
only applies to models with Bp--Rp $>$ 5 mag. 
	Synthetic isochrones for 47 Tuc (Zhou \etal 2022)
that do $not$ fit the cluster main sequence
reach 22nd absolute mag at intermediate colors when
an Eddington boundary condition is adopted.
\subsection{Captive PBHs}
Main sequence stars that have consumed a PBH also have smaller radii (Oncins \etal 2022). These models have
an m12 (10$^{-12} $ M$_\odot$) in their core. If m12 make up a fraction f of the local DM, their  frequency of entering a 1 AU cube containing a host star is 25 f v$_{100}$ /Myr, where v$_{100}$ is the DM velocity dispersion	in 100 \kms units. The dynamics of
capture and the hydrodynamics of immersion are worked out by these authors.
For a 0.32 M$_\odot$ star 10 Gyr after capture,
and for a 1 M$_\odot$ star 1.3 Gyr after capture,
Oncins \etal (2022) show  radius reductions
of 20--30\% from their zero age main sequence values. Models with larger
captive PBHs would be a useful addition.
 \subsection{Where are the PBHs?}
The Bp--Rp color range (0, 6) mag contains PBHs from 10$^{-8}$ to 10$^{-11}$ M$_\odot$. 
However, naked PBHs are so faint they lie 50 mag below $Gaia's$ limiting magnitude.
An upper limit can be placed on the number of m9 and m10 PBHs
present in the solar neighborhood.
PBHs detectable by $Gaia$ come within 0.01 AU, based on the absolute magnitudes in Figure 1. If they represent 100\% of the DM, their arrival rate is 0.25 $\times$ 10$^{-10}$ per year for a velocity of 100 \kms. This is a negligible rate compared with similarly close encounters with several asteroids per year\footnote{https://cneos.jpl.nasa.gov/} with their typical velocity of 30 \kms. Even the Rubin telescope with its ten times larger range will be unable to secure direct detection of PBHs. For Rubin microlensing is the best route to detecting asteroid mass PBH (Dent, Dutta, \& Xu 2025). 
PBHs of mass
m13--m15 fall orders of magnitude below $Chandra's$ limiting flux at keV energies
of 9 $\times$ 10$^{16}$ ergs cm$^{-2}~ s^{-1}$. Limits on 10$^{-20}$ M$_\odot$ PBH
have been placed by the Fermi-LAT collaboration (2018), effectively a fraction f of the local DM
in its Hawking radiation range of 100 MeV to 100 GeV ($\sim$m19--m22), f $<$ 0.2 with 99\% confidence.
\section{Conclusions}
Based on the collected evidence and models,
the faintest high proper motion stars in the $Gaia$ archive with G $>$ 20 mag are
\begin{itemize}
	\item~ PBHs: nil
	\item~ FFPs: some of those with the reddest colors
	\item~ Population II: $\sim$1\%
	\item~ captive PBHs: the best candidates are those with Bp--Rp $<$ 5
		and G $>$ 20.
	\item~ UCDs with anomalous radii $\lesssim$ R$_J$.
\end{itemize}
It will not be straightforward to distinguish further between these possibilities, e.g. captive
PBH might mix helium from the core to the surface, but low mass stars
are too cool for He absorption lines, and helium 10830\AA~ in flare stars does not
readily yield an He abundance.
The globular cluster $\omega$~Cen may have a main sequence with helium
abundance Y $>$ 0.3 (Bedin \etal 2004). If the anomaly is in Y, rather
than Z, captive PBH could be the cause.
The effective radius of the Milky Way's globular
clusters is half the solar radius in the Galaxy.
So on average, their environment is richer
in DM than the Sun's.

\subsection*{References}
\noindent Bedin, L. \etal 2004, ApJ, 605, L125 \\
Bottaro, S., Caputo, A., \& Fiorillo, D. 2024, JCAP, 11, 015\\ 
Coleman, G. \& Derocco, W. 2025, MNRAS, 537, 2303\\
Dent, J., Dutta, B. \& Xu, T. 2025,
 Phys Lett B, 861, id.139254\\
Fermi-LAT collaboration 2018, ApJ, 857, 49, arxiv 1802.00100\\
Heyl, J. \etal 2017, ApJ, 850, 186\\
Lutz, T. \& Kelker, D. 1973, PASP, 85, 573\\
Mazzi, A. \etal 2024, MNRAS,527, 583\\
Mould, J. 2025, ApJ, 984, 59\\ 
Oncins, M. \etal 2022, MNRAS, 517, 28\\
Phillips, M. \etal 2020, A\&A, 637, A8\\
Niikura, H. \etal 2019, Nature Astronomy, 3, 524\\
Richer, H. \etal 2008, AJ, 135, 2141\\
Sarro, L. \etal 2023, A\&A, 669, 139\\
Scholtz, J. \& Unwin, J. 2019, PRL, 125.051103\\
Sumi, T. \etal 2023, AJ, 166, 108\\
Suzuki, D. \etal 2016, 
ApJ, 833, 145 \\
Zhou, T. \etal 2022, Res. Notes AAS, 6, 212\\
\\
\leftline{\it \large \sc Acknowledgements}
ARC 
Grant CE200100008, 
ESA 
$Gaia$ 
DPAC.
\end{document}